# A Resilient AWGR and Server Based PON Data Centre Architecture

**Randa A. Thabit, Taisir E.H. El-Gorashi, and Jaafar M.H. Elmirghani**
*School of Electronic and Electrical Engineering, University of Leeds, LS2 9JT, United Kingdom*

**ABSTRACT**

This paper studies the resilience of an AWGR and server based PON DCN architecture against link failure scenarios and proposes a modified design for improved resilience. A MILP model is developed to evaluate the performance of the modified design considering different failure scenarios. The results show a limited increase in power consumption and a large increase in delay under failure scenarios compared to the normal state.

## 1. INTRODUCTION

The dramatic increase in Internet-connected devices and the growing popularity of data intensive applications over the last decade has created an increasing interest in designing energy-efficient communication networks. The work in [1] – [5] considered the energy efficiency of data centre networks and content distribution networks. Energy efficient virtualization of data centre and network resources was considered in [6] – [8]. The recent increase in the number of types of Internet of Things (IoT) applications has resulted in the generation of large amounts of big data that has to be processed at the edge of the network to improve the combined energy efficiency of the network and data processing [9] – [12]. The energy efficiency of core and end to end networks was evaluated in detail in [13] – [20] while new techniques were developed for coded data routing to improve the energy efficiency of protection routes in [21] – [22]. For data centres, there is a need for scalable, high bandwidth, low power consumption data centre architectures to meet the increasing demand. Passive Optical Networks (PONs) with their proven performance in access networks can provide efficient solutions to support connectivity inside modern data centres. Recently, a number of novel PON architectures were proposed to manage the inter-rack and intra-rack communication in a data centre [23] – [26]. In [27], [28] we studied one of the PON designs proposed in [25] where Arrayed Waveguide Grating Routers (AWGRs) and servers are used to route traffic. In this paper, we focus on the resilience of the AWGR and server based PON data centre architecture against different link failure scenarios and propose a modified design for improved resilience. A MILP model is developed to optimise traffic routing under different failure scenarios. The performance is evaluated in terms of power consumption and delay.

The reminder of this paper is structured as follows: In Section 2, the AWGR and server based-PON DCN architecture is introduced and Section 3 proposes a modified design for improved resilience. In Section 4, the power consumption and delay results are presented. Finally, conclusions are provided in Section 5.

## 2. THE AWGR AND SERVER BASED-PON DCN ARCHITECTURE

In this architecture, as shown in Fig. 1, the servers in each rack are placed in one or more groups where a group is composed of a number of subgroups. A subgroup can host a number of servers and is supported by a TDM PON whose splitting ratio defines the subgroup. The subgroups in every group are connected to a special server that acts as a gateway that provides inter-subgroup and inter-group communication. Each special server is connected to two AWGRs: one for sending to other groups and the other for receiving from them. The intra-subgroup communication is achieved using a fibre Bragg grating (FBG), which reflects only the wavelength assigned for the intra-subgroup communication. For the inter-subgroup communication and inter-group communication, the wavelength passes the FBG to the special server that route it to its destination.

## 3. RESILIENT AWGR AND SERVER BASED-PON DCN ARCHITECTURE

The main sources of failures in communication networks are link cuts, wear out, overload, malicious attacks and environmental disasters [29], [30]. To achieve enhanced network resilience, different techniques are used such as prevention, protection and restoration [21], [22], [29], [31].

In the AWGR and server based-PON DCN, we consider the failure of backplanes (which contain active components) and fibre link disruptions (due to transceiver failures). Table 1 gives the possible backplane and link failures that can affect the architecture. For the design in Figure 1, any fault in wiring or FBG will affect both inter subgroup and intra subgroup traffic. This architecture can survive only two out of eight possible types of link failures as illustrated in Table 1. Accordingly, for failure scenario S8, the traffic will use the other PON groups as a relay to reach its destination. Also, the architecture can survive S5 by using another normal server in the same subgroup as a relay then the traffic continues to the special server and back to its destination. All other kinds of link failure will affect the inter or intra subgroup communication or both.

For improved resilience the architecture is modified so the backplane provides connectivity at the rack level, i.e. the servers in each rack are connected by a backplane as shown in Fig.2. In this design, there are two types of communication: intra rack communication and inter rack communication. Intra rack traffic is separated from inter rack traffic where intra rack traffic passes through the backplane and inter rack traffic traverses through the special server. This connection makes the AWGR and server based PON data centre architecture resilient since it survives all the 8 possible types of link failure in Table 1. Under S1, the architecture can still allow communication between the affected server and servers in the same subgroup by using multi hop routing through other subgroups. Under S2, the server affected can reach another server in the subgroup through the backplane and can use it as a relay to reach the TDM PON coupler for inter subgroup communication. Under S3, the server affected can reach another server in the group through the backplane and use it as a relay to reach the other TDM PON coupler in that group for inter group communication. As for S5, the splitter can overcome it by sending the traffic to a server in the same subgroup as a relay then the traffic passes through the backplane to reach its destination. Furthermore, under S4, S6 or S7, the traffic uses the other special server in that rack as a relay to reach its destination while under S8, the traffic uses the other PON groups as a relay to reach its destination.

## 4. EVALUATION OF THE RESILIENT AWGR AND SERVER BASED PON DATA CENTRE ARCHITECTURE

A MILP model is developed to optimise traffic routing so that the total power consumption is minimized under different failure scenarios. The total power consumption includes the power consumed by the servers, special servers and OLT ports. This model is subject to a number of constraints to ensure that the total traffic of a server does not exceed its data rate, the total traffic of a special server does not exceed its fitted ONU data rate, and the total traffic of an OLT port does not exceed its data rate. This model is used to minimize the power consumed by servers, special servers and OLT ports.

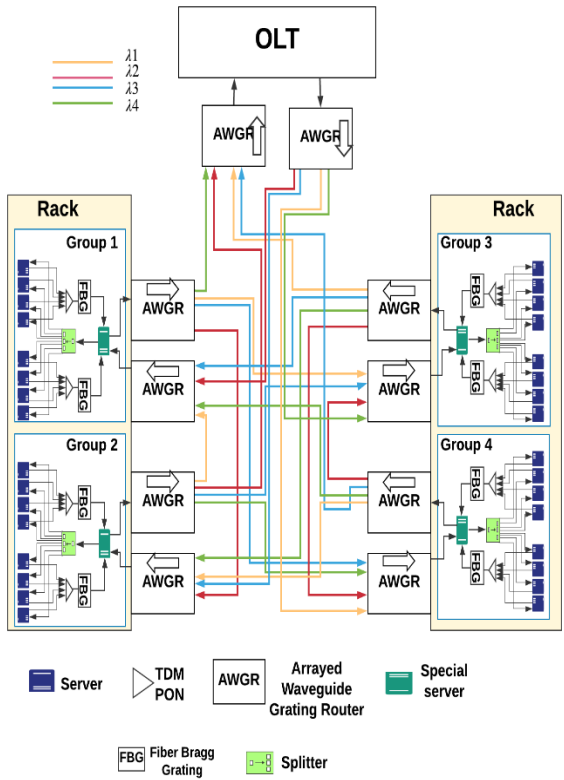
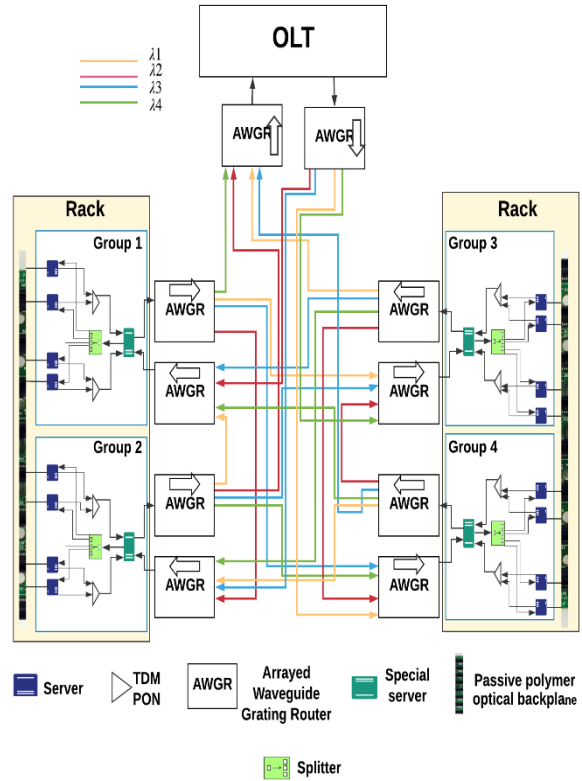

*Figure 1: The AWGR and server based PON data centre architecture*

*Figure 2: The modified design of AWGR and server based PON data centre architecture*

*Table 1: Resilience of a AWGR and server based PON data centre architecture and the modified design against link failure*

| Architecture | Link failure | S1 | S2 | S3 | S4 | S5 | S6 | S7 | S8 | S9 | S10 |
|---|---|---|---|---|---|---|---|---|---|---|---|
| | | Backplane to a server link | Server to TDM PON link | Link between TDMPON and special server | Link from special server to Splitter | Splitter link to server | Link from special server to AWGR | Link from AWGR to special server | Link between AWGRs | Link between TDMPON and FBG | Link between FBG and special server |
| The AWGR and server based PON data centre | | NA | No | NA | No | Yes | No | No | Yes | No | No |
| The modified design | | Yes | Yes | Yes | Yes | Yes | Yes | Yes | Yes | NA | NA |

*Table 2: Input data for the model*

| Parameter | Value |
|---|---|
| Traffic demand between servers $TD_{sd}$ | 200-800 Mbps random and uniformly distributed |
| Capacity of physical link $L_{mn}$ | 10 Gbps |
| Large enough number $M$ | 100000 |
| Idle power consumption of a server $PSI$ | 301.6 W [31] |
| Maximum power consumption of a server $PSM$ | 457 W [31] |
| Idle power consumption of an OLT port $OIP$. | 2 W |
| Maximum operational power consumption of an OLT port. | 14.3 |
| Fraction of a server's processing capacity used for forwarding one traffic demand. | 1.5% |
| Fraction of a server's processing capacity used for transmitting one traffic demand | 0.3% |
| Fraction of a server's processing capacity used for receiving one traffic demand | 0.2% |
| Server's Data rate | 1 Gbps |
| OLT's port data rate | 10 Gbps |
| Total ONU power consumption | 2.5 W [32] |
| ONU data rate | 10 Gbps |

We evaluate the performance of a data centre cell of 16 servers placed into two racks each hosting 8 servers. Each rack contains two groups which are further divided into subgroups. Servers in each rack are connected by a backplane as depicted in Fig.2. Table 1 presents the studied failure scenarios while Table 2 presents the input parameters used for the model.

Fig. 3 shows the power consumption of the modified design in the normal state (no failure (NF)) and under failure scenarios. Regarding link failure S1, the demands of the affected server will communicate with the special server and back to its destination. This will add special servers power consumption for forwarding demands which is considerably small. This results in a slight difference in power consumption of 0.2% compared to the normal state. For S2, S3, S4, S5, S6 and S7, only the demands affected by these link failures need to hop through one more server which adds power consumption due to the forwarding server power consumption which is small. This explains the slight difference in power consumption of 0.6%, 1.9%, 1.3%, 1.5%, 1.3% and 1.3% respectively compared to the normal state. Regarding the link failure S8, a relay PON group is used to recover from this failure which means traversing through one more special server. Again, this will add special servers power consumption for forwarding demands which is considerably small of 0.3% compared to the normal state.

Fig. 4 shows the queuing delay of the modified design in the normal state and under failure scenarios. For link failure S1, only the affected server demands need to be routed through the special server and back to their destination. This causes the queuing delay to increase by 2% compared to the normal state. As for link failures S2 and S5, only the affected server demands need to be relayed through a server which increases the queuing delay by 51% compared to the normal state. For S3, the demands of the affected subgroup need to traverse through the backplane to be relayed by other servers in the same rack. This increases the queuing delay by 49% compared to the normal state. As for S4, S6 and S7, these failures influence all the inter group demands transmitted and destined to the affected group of servers, which causes them to be relayed by the other servers and a special server in their rack to reach their destination. This in turn causes considerable difference in queuing delay of 131%, 87% and 65% respectively compared to the normal state. As for link failure S8, the affected demands need to be relayed through only a special server. The relaying special server is selected so minimum additional power is consumed, i.e. an already activated special server is preferred. Here, in Fig. 4, the special server with lower power consumption used to relay the traffic is the one in the same rack of the affected special server. The relaying continues through relay server(s) using the backplane to reach the destination. This is why there is considerably higher queuing delay of 35% more than the normal state.

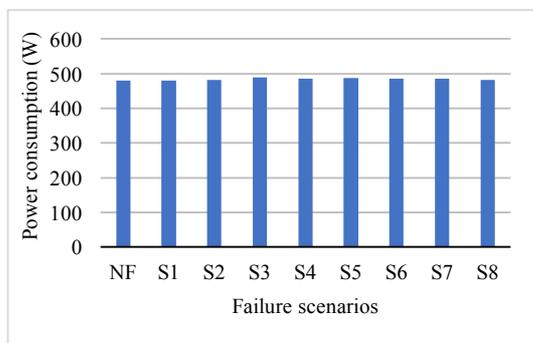

*Figure 3: Power consumption for the modified architecture under different failure scenarios*

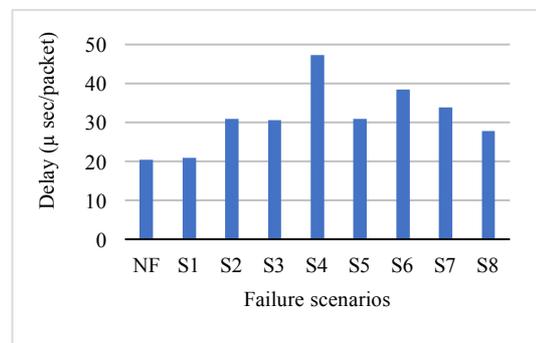

*Figure 4: Delay for the modified architecture under different failure scenarios*

## 5. CONCLUSIONS

In this paper, we investigated the resilience of the AWGR and server based PON data centre architecture against different link failure scenarios and proposed modified designs for improved resilience. A MILP model was developed to optimize traffic routing under different failure scenarios. The results show limited increase in power consumption by up to 1.9% under failures compared to the normal state. However, there is significant increase in delay in some failure scenarios up to 131% compared to the normal state.

**ACKNOWLEDGEMENTS**

The authors would like to acknowledge funding from the Engineering and Physical Sciences Research Council (EPSRC), INTERNET (EP/H040536/1), STAR (EP/K016873/1) and TOWS (EP/S016570/1). Mrs. Randa A. Thabit would like to acknowledge The Iraqi Ministry of Higher Education and Scientific Research for funding her scholarship. All data are provided in full in the results section of this paper.